\documentclass[]{raa}
\usepackage{graphicx,times}
\input{epsf.sty}
\input{psfig.sty}

\newcommand{\mt}{\hbox{$M_{\rm 20}$}}
\newcommand{\etal}{\hbox{et al.\,}}
\newcommand{\Kv}  {\hbox{$K_{\rm Vega}$}}
\newcommand{\gsim}{\lower.5ex\hbox{$\; \buildrel > \over \sim \;$}}
\newcommand{\bsex}{\hbox{\tt SExtractor}}
\newcommand{\zph}  {\hbox{$z_{\rm phot}$}}
\newcommand{\zsp}  {\hbox{$z_{\rm spec}$}}

\begin{document}

   \title{Classification of Extremely Red Objects in the Hubble
Ultra Deep Field }

   \volnopage{Vol.0 (200x) No.0, 000--000}
   \setcounter{page}{1}

   \author{Guan-Wen Fang
      \inst{1}
   \and Xu Kong
      \inst{1,2}
   \and Min Wang
      \inst{1}
      }

   \institute{Center for Astrophysics, University of Science
and Technology of China, Hefei, Anhui, 230026, China;\\
        \and Joint Institute for Galaxy and Cosmology
(JOINGC) of SHAO and USTC;  {\it xkong@ustc.edu.cn}\\
          }

   \date{Received~~2008 month day; accepted~~2009~~month day}

\abstract{
In this paper we present a quantitative study of the classification
of Extremely Red Objects (EROs).
The analysis is based on the multi-band spatial- and ground-based
observations (HST/ACS-$BViz$, HST/NICMOS-$JH$, VLT-$JHK$) in the 
Hubble Ultra Deep Field (UDF). 
Over a total sky area of 5.50 arcmin$^2$ in the UDF, we select 24 
EROs with the color criterion $(i-K)_{\rm Vega}>3.9$, corresponding 
to $(I-K)_{\rm Vega}\gsim4.0$, down to $\Kv=22$.
We develop four methods to classify EROs into Old passively evolving
Galaxies (OGs) and Dusty star-forming Galaxies (DGs), including
$(i-K)$ vs. $(J-K)$ color diagram, spectral energy distribution
fitting method, Spitzer MIPS 24 $\mu$m image matching, and
nonparametric measure of galaxy morphology, and found that the
classification results from these methods agree well. Using these
four classification methods, we classify our EROs sample into 6 OGs
and 8 DGs to $\Kv<20.5$, and 8 OGs and 16 DGs to $\Kv<22$,
respectively. The fraction of DGs increases from 8/14 at $\Kv<20.5$
to 16/24 at  $\Kv<22$.
To study the morphology of galaxies with its wavelength, we measure
the central concentration and the Gini coefficient for the 24 EROs
in our sample in HST/ACS-$i,z$ and HST/NICMOS-$J,H$ bands.
We find that the morphological parameters of galaxies in our sample
depend on the wavelength of observation, which suggests that caution
is necessary when comparing single wavelength band images of 
galaxies at a variety of redshifts.
 }

   \authorrunning{Fang \etal }
   \titlerunning{EROs in the UDF }

   \maketitle


\section{Introduction}
\label{sect:intro}

Extremely Red Objects (EROs) are massive galaxies ($M_\ast>10^{11}
M_\odot$), characterized by extremely red optical-to-infrared colors
and high redshifts (Hu \& Ridgway~\cite{hu94}; Elston et
al.~\cite{elst88}, \cite{elst89}; Stern et al.~\cite{ster06};
Conselice et al.~\cite{cons08a}). EROs are now instead recognized to be
primarily comprised of two interesting galaxy populations: {\bf O}ld
passively evolving {\bf G}alaxies (hereafter {\bf OGs})
characterized by old stellar populations, and {\bf D}usty
star-forming {\bf G}alaxies (hereafter {\bf DGs}) reddened by a
large amount of dust.  EROs continue to attract considerable
interest, on the one hand, the research in the literature suggests
that they may be the direct progenitors of present-day massive E/S0
galaxies. On the other hand, they can provide crucial constraints
on the current galaxy formation and evolution models (Kitzbichler \&
White~\cite{kitz07}). Therefore, the key question is
then to measure the relative fraction of both galaxy types in order
to exploit the stringent clues that EROs can place on the formation
and evolution of elliptical galaxies and the abundance of dust
obscured system at high redshift.

Many groups are currently investigating the fractions of these two
ERO populations using a variety of observational approaches, but the
fraction of OGs and DGs from different surveys is different.  Some
work found that OGs were dominant in EROs (Moriondo et
al.~\cite{mori00}; Simpson et al.~\cite{simp06}; Conselice et
al.~\cite{cons08a}), but the others reported nearly the opposite
results, and found that most of EROs with spiral-like or irregular
morphology (Yan \& Thompson.~\cite{yan03}; Cimatti et
al.~\cite{cima03}; Sawicki et al.~\cite{sawi05}). In addition, some
authors also reported that the OGs and DGs have similar fractions in
their EROs sample (e.g. Mannucci et al.~\cite{mann02}; Giavalisco et
al.~\cite{giav04}; Moustakas et al.~\cite{mous04}). Therefore, one
of the main open questions about EROs is the relative fraction of
different ERO types.

To determine the relative fraction of different EROs accurately, we
develop four methods for DGs and OGs classification, such as the
$(i-K)$ vs. $(J-K)$ color diagram, the multi-wavelength spectral
energy distribution (SED) fitting method, the Spitzer MIPS
(Multiband Imaging Photometer for Spitzer) 24 $\mu$m image matching
method, and the nonparametric measures of galaxy morphology method,
including Gini coefficient ($G$), the second-order moment of the
brightest 20\% of the galaxy's flux (\mt), and rotational asymmetry
($A$) (Abraham et al.~\cite{abra96}, \cite{abra03}; Bershady et
al.~\cite{bers00}; Conselice~\cite{cons03}; Lotz et
al.~\cite{lotz04}, \cite{lotz08}; Conselice et al.~\cite{cons08b}).
To check the reliability of these methods, for the first time, we
applied our methods to the EROs sample over the Hubble Ultra Deep
Field (hereafter UDF) in this paper. We will apply these methods for
large data sets, such as GEMS and COSMOS in the future (Rix et
al.~\cite{rix04}; Scoville et al.~\cite{scov07}).

We describe the multi-band spatial- and ground-based observations of
the UDF; introduce data reduction and method for obtaining our EROs
sample in Section 2.  Section 3 presents the four classification
methods of EROs and their application for EROs in the UDF. We
present classification result and discuss the morphological 
parameters of EROs varying with wavelength in Section 4, and summarize 
our conclusions in Section 5.
Throughout this paper, all magnitudes and colors are in the Vega 
system unless stated otherwise\footnote{The relevant
conversion between AB and Vega magnitudes for this paper are
$K_{\rm AB}=\Kv+1.87$, 
$i_{\rm AB}=i_{\rm Vega}+0.41$.
}.

\section{Observations, Data Reduction and EROs Selection}
\label{sect:obs}

\subsection{Observations}

The UDF field lies within the $Chandra$ Deep Field South (CDF-S, or
Great Observatories Origins Deep Survey South, GOODS-S) with
coordinates RA = 03$\rm ^h$32$\rm ^m$39.0$\rm ^s$, Dec =
-27$^\circ$47$'$29.1$''$ (J2000) (Giavalisco et al.~\cite{giav04};
Beckwith et al.~\cite{beck06}). The field has been imaged by a
large number of telescopes at a variety of wavelengths (Coe et
al.~\cite{coe06}). In this
paper, HST/ACS (Advanced Camera for Survey) images, HST/NICMOS
(Near-Infrared Camera and Multi-Object Spectrometer) images,
VLT/ISAAC (Infrared Spectrometer And Array Camera) images and a 
Spitzer/MIPS 24 $\mu$m image of the UDF were used.

With a total of 544 orbits, the UDF is one of the largest time
allocations of HST, and indeed the filter coverage, depth, and
exquisite quality of the UDF ACS and NICMOS images provide an
unprecedented data set for the study of galaxy morphology, even of
very low surface brightness components.  They are taken in four
optical bands and two near-infrared bands: $B$(F435W), $V$(F606W),
$i$(F775W), $z$(F850LP), $J$(F110W) and $H$(F160W). Due to the small
field of NICMOS camera, the UDF NICMOS only covers a subsection
(5.76 arcmin$^2$) of the optical UDF (11.97 arcmin$^2$).  For our
analysis, we use the reduced UDF optical imaging data v1.0 made
public by STScI on 2004 March 9. The 10 $\sigma$ limiting AB
magnitudes are 28.7, 29.0, 29.0 and 28.4 for $B$-, $V$-, $i$- and
$z$-band (Beckwith et al.~\cite{beck06}). The $J$- and $H$-band
data are given by Thompson \etal (2005). The 5 $\sigma$ limiting AB
magnitudes is 27.7 at 1.1 and 1.6 $\mu$m in a 0.6$''$ diameter
aperture.

In addition to the HST NICMOS and ACS data, we also use the Spitzer MIPS
image. MIPS is one of the facilities on the Spitzer Space Telescope
that is used to image at 24, 70 and 160 $\mu$m. In this paper, we
use the super deep 24 $\mu$m image data only, which is part of the
GOODS Spitzer Legacy Survey (PI: Mark Dickinson).  GOODS/MIPS Data
Release v3 was used for our analysis.  In addition, ground-based
near-infrared images ($JHK$) of the UDF are taken as part of the
GOODS with VLT/ISAAC.  GOODS/ISAAC Data Release v1.5 was used for
our analysis. They are reduced using an improved version of the
ESO/MVM image processing pipeline. The 5 $\sigma$ limiting AB
magnitudes at $J$-, $H$- and $K$-band are 25.3, 24.8 and 24.4 in a
2.0$''$ diameter aperture.

\subsection{Data reduction}

\begin{figure*}
\centering
\includegraphics[angle=0,width=0.85\textwidth]{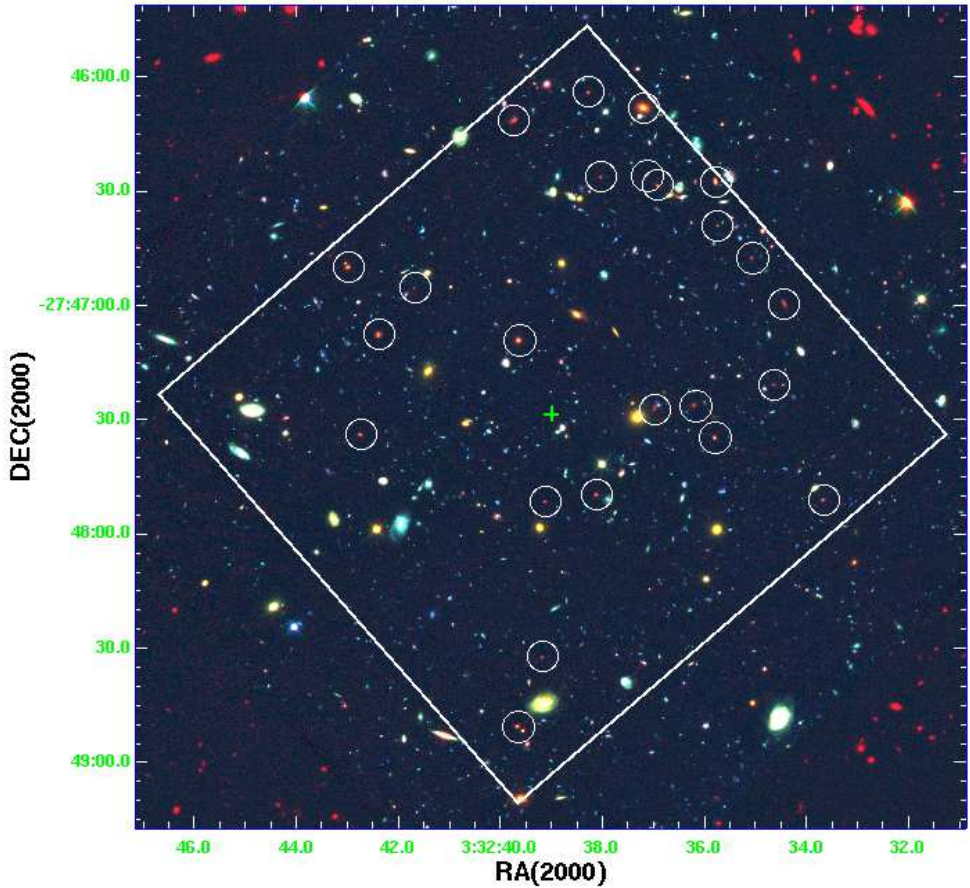}
\caption{Composite pseudo-color image of the UDF. The RGB colors are
assigned to VLT-$K$, ACS-$z$, and ACS-$B$ band images. The outlined
(white) region of the image is the field where the NICMOS-$JH$ band
images have a high signal-to-noise ratio (5.5 arcmin$^2$). The small
white circles show the sky positions of 24 EROs in the UDF.}
\label{fig:area}
\end{figure*}

All of these data have been released in fully processed form and no
additional processing is necessary. However, considering that the
images in different data sets have different scales and sizes, so
they must be resampled to put them on the same astrometric grid. The
resampling was done with IRAF's $geomap$ and $geotran$ tasks.  All
images were remapped to 0.09$''$ pixel$^{-1}$, the same scale as the
NICMOS $J$-band and $H$-band images. By comparing the resampled
images around the galaxy luminosity, we found that the resampling
process of the sample could not cause each source deviation in the
position or the flux. In addition, the edges of the HST/NICMOS images
have only one integration, as compared to the average 16
integrations for the interior of the images.  Therefore, the edges
of the resulting final images, where the signal-to-noise of
HST/NICMOS images was very low, were then trimmed.
Figure~\ref{fig:area} shows a composite pseudo-color image of the
UDF. The area, as discussed in this paper, was reduced from
HST/NICMOS's 5.76 to 5.50 arcmin$^{2}$, as the white outlined region 
of the image.

Source extraction in the science image was performed with the
program \bsex\ version 2.5 (Bertin \& Arnout.~\cite{bert96}) in the
dual image mode, a 2$''$ diameter aperture was used
for aperture magnitudes.  More detailed description about catalog
construction can be found in Kong et al.(~\cite{kong08a}). Compared
to the optical selection, the near-infrared selection has several
advantages, in particular in the K-band (Broadhurst et
al.~\cite{broa92}; Kauffmann \& Charlot.~\cite{kauf98}). Therefore,
we select objects to $K<22$ over a total sky area of
5.5\,arcmin$^2$ in the UDF, and 210 objects (including 202 galaxies
and 8 stars, star-galaxy separation using the same method as that in
Kong \etal 2006) were included in our final catalog.
A comparison of the $K$-band number counts in the UDF survey with a
compilation of counts published in the literature can be found in
the figure 2 of Kong et al. (2008a).
The red-, black-, green- and blue-filled
squares correspond to the counts of field galaxies in the UDF (this
paper), COSMOS (Kong et al.~\cite{kong08b}), Daddi-F and Deep3a-F
(Kong et al.~\cite{kong06}), respectively.  As shown in that figure,
our number counts in different fields are in good agreement with
those of the previous surveys.

\subsection{ERO Sample Selection}

\begin{figure*}
\centering
\includegraphics[angle=-90,width=0.85\textwidth]{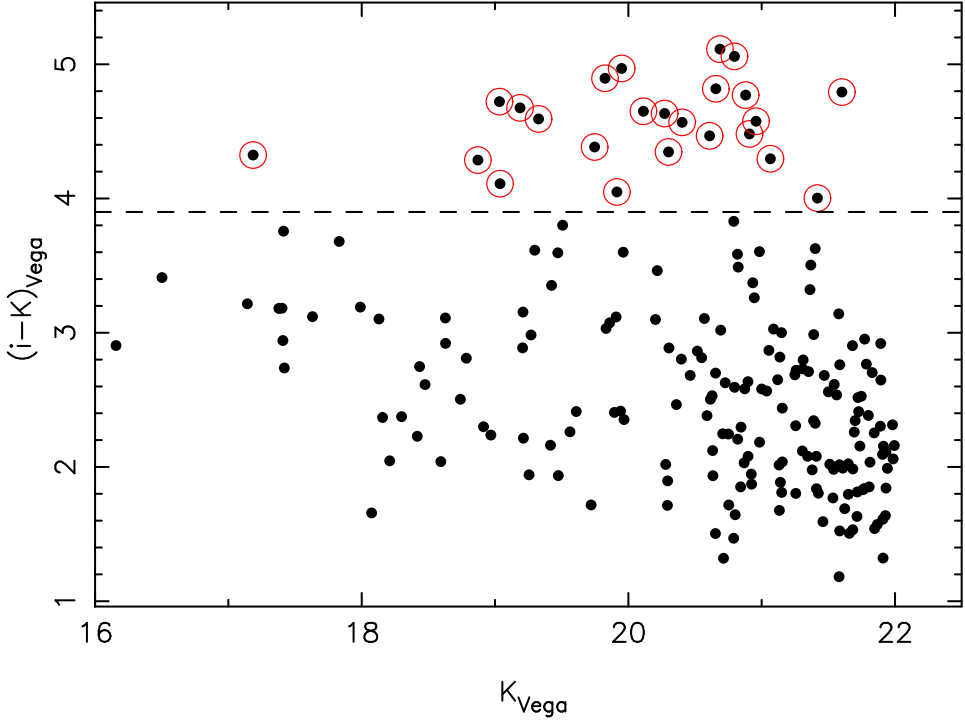}
\caption{$K$-band magnitude vs. $(i-K)$. All galaxies in
the $K$-limited sample in the UDF are plotted as filled circles, and
EROs are overlaid with larger open circles.
The dashed line shows the threshold of
$(i-K)$ = 3.9 for selecting EROs.}
\label{fig:cmd}
\end{figure*}

Numerous different selection criteria have been defined for EROs,
including $R-K\geq 6$, $R-K\geq 5.3$, $R-K\geq5$, $I-K\geq 4$ in the
Vega magnitude system with $K-$magnitude upper limits from 18 to 20,
or $R-$[3.6] $\geq 4$ in the AB magnitude system (e.g. Hu \&
Ridgway.~\cite{hu94}; Scodeggio \& Silva.~\cite{scod00}; Brown et
al.~\cite{brow05}; Kong et al.~\cite{kong06};
D\'{\i}az-S$\acute{a}$nchez et al.~\cite{diaz07}; Wilson et
al.~\cite{wils07}). Since the UDF field has not $I-$band or $R-$band
observation data, we cannot use $R-K$ or $I-K$ color criteria for
EROs selection. In this paper, we use ACS-$i$ and ISAAC-$K$ for EROs
selection. We calculated $i-K$ color using SEDs from the Kodama \&
Arimoto (1997, KA97)'s library, and found $(i-K) > 3.9$
can be used to select both ellipticals and reddened starbursts when
their redshift is beyond 0.8. Therefore, EROs in this paper are
selected by $(i-K) > 3.9$, which corresponds to
$(I-K) > 4$. 2$''$ diameter aperture magnitudes were
used for color calculation.

Figure~\ref{fig:cmd} shows the color-magnitude plot for all 202
galaxies, with $K<22$, in the UDF. 24 galaxies with $(i-K) > 3.9$ 
were selected as EROs, and they are plotted in
Figure 2 as larger circles, for a surface density of 4.36
arcmin$^{-2}$ to $K<22$. The $K$-band differential number counts
of EROs in the UDF are plotted in the figure 2 of Kong et al. (2008a)
also, as those of field galaxies.
We found that the differential number
counts of EROs in different fields are in good agreement. In
addition, the slope of the number counts of EROs is a variable, being
steeper at bright magnitudes and flattening out towards faint
magnitudes. A break in the counts is present at $K\sim18$, very
similar to the break in the ERO number counts observed by previous
works.

\section{Classification of EROs}\label{sec:cls}

Many researchers have attempted to separate the
two types of EROs by various methods. However, their results
conflict with each other. In order to estimate the relative fraction
of OGs and DGs in EROs, we develop four classification methods, and
try to apply them to the ERO sample of the UDF in this section.

\subsection{Classification based on SED Fitting}\label{sec:sed}

\begin{figure*}
\centering
\includegraphics[angle=-90,width=0.85\textwidth]{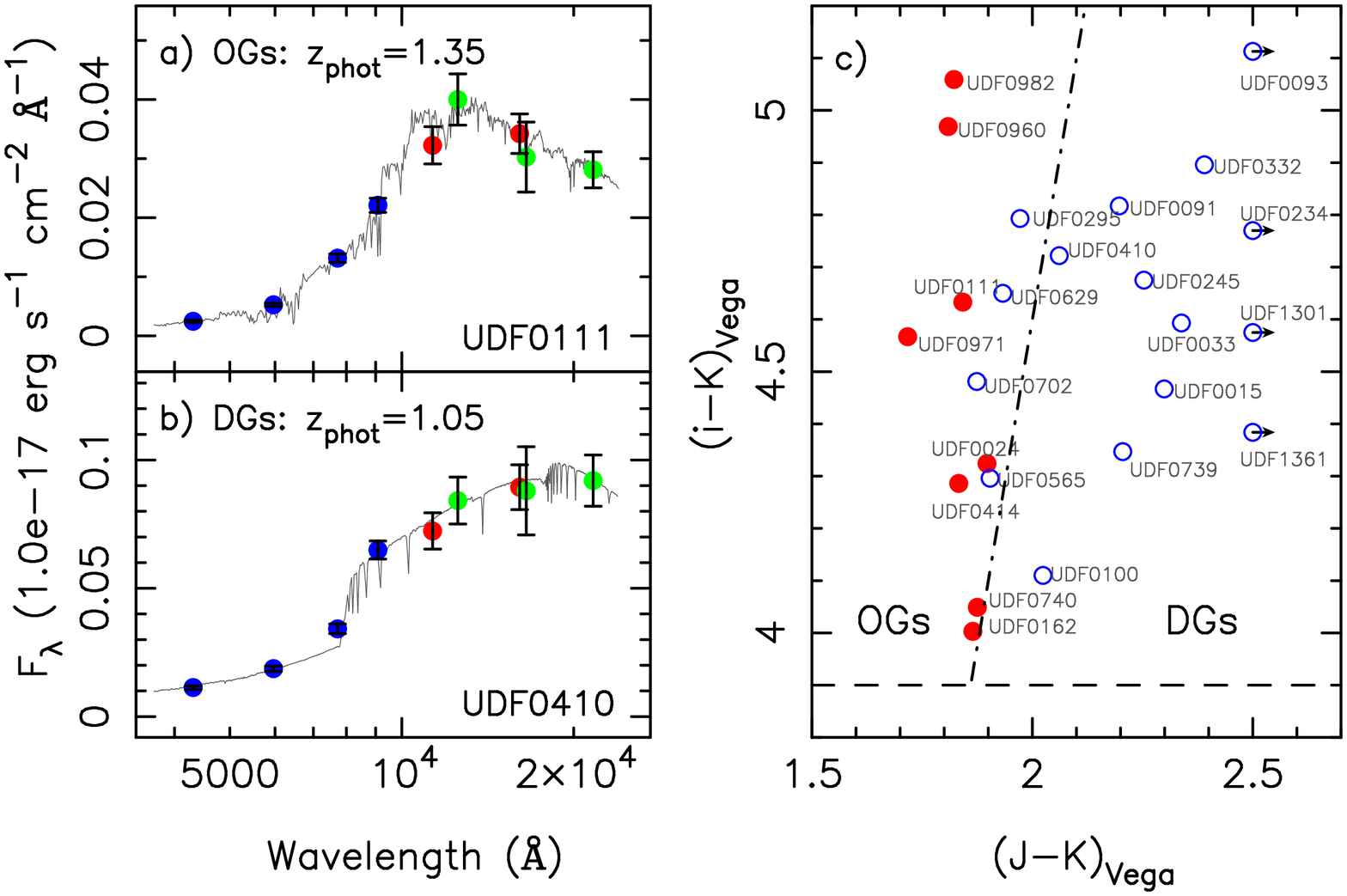}
\caption{Left panel: Spectral energy distribution of two EROs in the
UDF. Filled circles show the observed photometric data, and solid
curves show the best fitting templates. Photometric redshift and
name for each ERO are shown also. Blue, red and green colors are
assigned to ACS$-BViz$, NICMOS$-JH$ and VLT-$JHK$ bands,
respectively.  
Right panel: Distribution of EROs on $(i-K)$ versus $(J-K)$ diagram. 
EROs classified as OGs by the SED fitting method are plotted in 
red filled circles, and as those DGs are plotted as blue open circles. 
The dashed line correspond to the threshold of $(i-K) = 3.9$ for
selecting EROs in this paper, the dot-dashed line is a new color
criterion for EROs classification, based on evolutionary population
synthesis model (see the next subsection for detail).}
\label{fig:csed}
\end{figure*}

An SED fitting technique based on the photometric redshift code
$HyperZ$ (Bolzonella et al.~\cite{bolz00}) is used to classify 
our ERO sample into different types, dusty and evolved,
 using their multi-waveband photometric properties.
The efficiency of the method is based on the fit of the overall
shape of the spectra and the detection of strong spectral features,
such as the 4000{\AA} break, the Balmer break, or strong emission lines
(Smail \etal 2002; Miyazaki \etal 2003; Georgakakis \etal 2008;
Stutz \etal 2008).

To illustrate this point more clearly, we show the SEDs of 2 EROs,
as an example, in the left panel of Figure~\ref{fig:csed}.  The
blue-, red-, and green-filled circles correspond to the observed
data of ACS-$BViz$, NICMOS-$JH$ and ISAAC-$JHK$ bands, respectively.
The best fitting SEDs of these two galaxies using the  photometric
redshift technique are shown as grey lines. The upper-panel
represents typical OGs, and the lower panel represents a typical
DGs. From this figure, we can find that OG shows a large break at
4000{\AA} rest-frame and the flux decreases at near-infrared band;
DGs shows a small 4000{\AA} break, a flat SED at optical and
near-infrared rest-frame wavelength, because of dust extinction and
dust radiation. As a result, we can use the SED fitting method to
classify EROs into OGs and DGs.

\begin{figure*}
\centering
\includegraphics[angle=-90,width=0.85\textwidth]{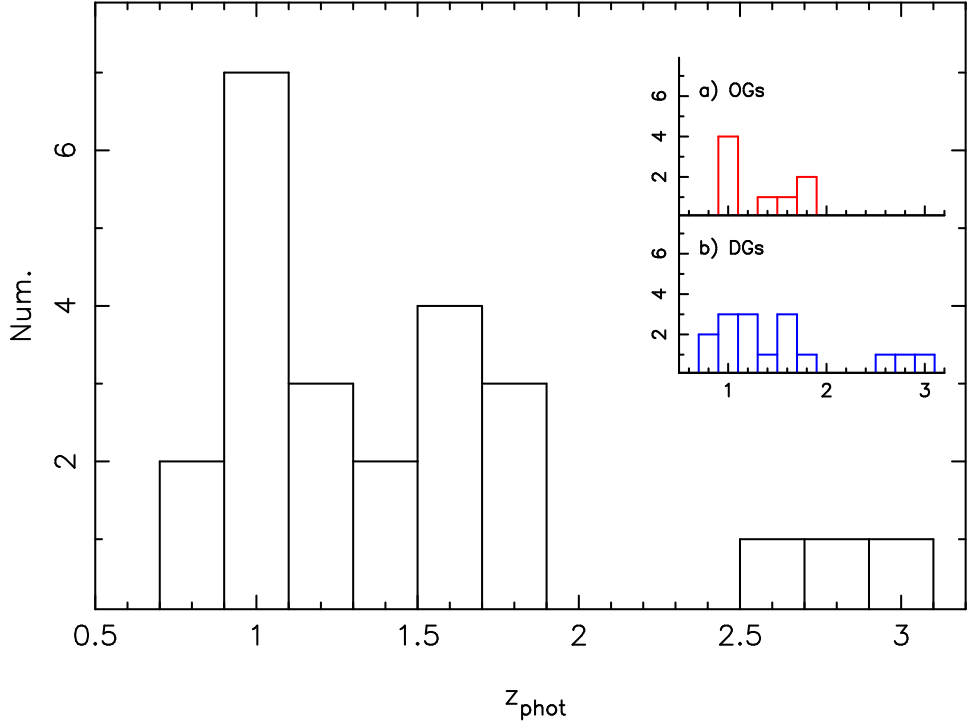}
\caption{
Photometric redshift distribution of EROs in the UDF. Panel a) 
shows the redshift distribution of OGs, and Panel b) shows the 
redshift distribution of DGs.
}
\label{fig:zphot}
\end{figure*}

We use a stellar population synthesis model by KA97 to make template
SEDs, similar to Miyazaki et al.(~\cite{miya03}).  KA97 includes the
chemical evolution of gas and stellar populations, and have been
successfully used to obtain photometric redshifts of high and low
redshift galaxies (Kodama et al.~\cite{koda99}, Furusawa et
al.~\cite{furu00}). 
The template SEDs consist of the spectra of pure disks, pure bulges, 
and intermediate SED types. Pure disk SEDs correspond to young or 
active star-forming galaxies, and pure bulge SEDs correspond to 
elliptical galaxies. The intermediate SED types (composites) are 
made by combining a pure bulge-like spectrum and a pure disk-like 
spectrum, the ratio of the bulge luminosity to the total luminosity 
in the $B$ band is changed from 0.1 to 0.99.
Full details of the templates can be found in Furusawa et 
al.(~\cite{furu00}) and Kong et al.(~\cite{kong08b}). 
The SED derived from the observed magnitudes of each object, 
including ACS-$BViz$, NICMOS-$JH$, and ISAAC-$JHK$, is compared to 
each template spectrum (redshift from 0.0 to 5.0 with step 0.05; 
$A_{\rm V}$ from 0.0 to 6.0 with step 0.05; internal reddening law 
introduced by Calzetti \etal 2000) in turn.
Figure 3 in Kong et al.(~\cite{kong08a}) shows a comparison of the
photometric redshifts from our SED fitting method with their
spectroscopic redshifts for 33 galaxies in the UDF.  From that
figure, we found that our photometric redshifts fit the
spectroscopic redshifts well, with an average $\delta z/(1+\zsp)
=0.02$. 

Figure~\ref{fig:zphot} shows the photometric redshift histogram of 
EROs in our sample. The peak of the redshift distribution is at 
$\zph \sim 1.0$, all of them have $\zph>0.8$. 
Only 3 EROs have redshift $\zph>2.0$, those are faint dusty 
EROs. 
The small panels in Figure~\ref{fig:zphot} show the redshift 
distribution of OGs (in panel a) and DGs (in panel b).  It is worth 
noting that OGs have a narrow redshift distribution, none of them 
have redshift higher than 2.0.
Photometric redshifts (\zph) and absorption in the $V$ band 
($A_{\rm V}$) of EROs are listed in Table~\ref{tab:cls} also.

Based on this SED fitting method, we classify EROs as DGs, if the
best fitting template of it is the spectra of disk-dominant (late
type SEDs); the others were classified as OGs (early type SEDs). 
Out of the 24 EROs in our sample, 8 are classified as OGs, while 
16 are classified as DGs. The results are listed in 
Table~\ref{tab:cls}.  Column (1) lists the galaxy name. 
Columns (2) and (3) list the right ascension and declination at
epoch 2000; units of right ascension and declination are degree. 
Column (4) lists $K$-band total magnitude in the Vega system; 
Columns (5) and (6) list the color of $i-K$ and $J-K$ in the
Vega magnitude. 
Columns (7) and (8) list the dust extinction ($A_{\rm V}$) and 
photometric redshift (\zph) for each source. 
Columns (9) -- (12) list the morphological parameters of 
concentration index ($C$), Gini coefficient, \mt, and rotational 
asymmetry (see Section~\ref{sec:mor}). 
Column (13) lists the classification results of EROs with the SED 
fitting method (M1, the first classification method in this paper).

\begin{table}
\setlength{\tabcolsep}{1.50mm}
\begin{center}
\caption[]{Properties, Nonparametric Morphological
Indicators and Classification of EROs in the UDF.\label{tab:cls}}
{\scriptsize
\begin{tabular}{cccccccccccccccccccccccccccccc}
\noalign{\smallskip}
\hline
\hline
\noalign{\smallskip}
NAME&RA&DEC&$K_{\rm tot}$&i-K&J-K&$A_{\rm V}$&$z_{phot}$&$C$&$A$&$G$&\mt&M1&M2&M3&M4&End\\
(1)&(2)&(3)&(4)&(5)&(6)&(7)&(8)&(9)&(10)&(11)&(12)&(13)&(14)&(15)&(16)&(17)\\
\hline
\noalign{\smallskip}
UDF0015&53.1593971&-27.7677822&20.61&4.47&2.30&0.95&1.04&0.36&0.035&0.53&-1.450&DG&DG&DG&DG&DG\\
UDF0024&53.1549187&-27.7689114&17.18&4.32&1.90&0.10&1.04&0.58&0.023&0.68&-2.197&OG&OG&OG&OG&OG\\
UDF0033&53.1655121&-27.7698059&19.33&4.59&2.34&1.00&1.45&0.33&0.172&0.50&-1.725&DG&DG&DG&DG&DG\\
UDF0091&53.1546974&-27.7738743&20.66&4.82&2.20&1.15&0.84&0.17&0.010&0.37&-0.842&DG&DG&OG&DG&DG\\
UDF0093&53.1583939&-27.7739716&20.69&5.11&2.68&2.20&1.80&0.41&0.128&0.56&-1.738&DG&DG&DG&DG&DG\\
UDF0100&53.1490250&-27.7743092&19.04&4.11&2.02&2.85&0.92&0.31&0.121&0.52&-1.361&DG&DG&DG&DG&DG\\
UDF0111&53.1537590&-27.7745819&20.27&4.63&1.84&0.15&1.35&0.49&0.019&0.65&-1.657&OG&OG&OG&OG&OG\\
UDF0162&53.1488419&-27.7775211&21.42&4.00&1.87&0.80&1.82&0.52&0.122&0.68&-1.413&OG&OG&DG&OG&OG\\
UDF0234&53.1460953&-27.7798824&20.88&4.77&3.38&1.70&2.72&0.33&0.111&0.46&-1.516&DG&DG&DG&DG&DG\\
UDF0245&53.1790085&-27.7805386&19.19&4.68&2.25&1.30&0.89&0.37&0.171&0.55&-1.364&DG&DG&DG&DG&DG\\
UDF0295&53.1736145&-27.7820663&21.60&4.79&1.77&0.80&1.56&0.21&0.029&0.40&-0.607&DG&OG&DG&DG&DG\\
UDF0332&53.1434212&-27.7832375&19.83&4.90&2.39&2.00&1.14&0.18&0.077&0.41&-0.406&DG&DG&DG&DG&DG\\
UDF0410&53.1765213&-27.7854481&19.03&4.72&2.06&0.75&1.05&0.31&0.112&0.54&-1.172&DG&DG&DG&DG&DG\\
UDF0414&53.1651039&-27.7858753&18.87&4.29&1.83&0.55&0.94&0.45&0.054&0.60&-1.920&OG&OG&OG&OG&OG\\
UDF0565&53.1442566&-27.7891445&21.07&4.30&1.90&1.25&1.15&0.21&0.085&0.47&-1.163&DG&OG&DG&DG&DG\\
UDF0629&53.1507835&-27.7906094&20.11&4.65&1.93&3.60&1.51&0.38&0.082&0.58&-1.620&DG&OG&OG&DG&DG\\
UDF0702&53.1539879&-27.7908936&20.91&4.48&1.87&2.90&1.63&0.23&0.108&0.43&-1.145&DG&OG&DG&DG&DG\\
UDF0739&53.1780281&-27.7927475&20.30&4.35&2.21&1.05&1.24&0.14&0.148&0.32&-1.044&DG&DG&DG&DG&DG\\
UDF0740&53.1491013&-27.7929821&19.91&4.05&1.88&0.50&1.09&0.50&0.064&0.64&-1.806&OG&OG&OG&OG&OG\\
UDF0960&53.1587563&-27.7971535&19.95&4.97&1.81&0.05&0.93&0.49&0.030&0.68&-1.542&OG&OG&OG&OG&OG\\
UDF0971&53.1402855&-27.7975273&20.40&4.57&1.72&0.80&1.50&0.50&0.077&0.64&-1.749&OG&OG&OG&OG&OG\\
UDF0982&53.1629448&-27.7976551&20.80&5.06&1.82&0.25&1.90&0.36&0.067&0.60&-1.424&OG&OG&OG&OG&OG\\
UDF1301&53.1631813&-27.8089867&20.96&4.58&3.35&1.70&2.63&0.31&0.077&0.58&-1.362&DG&DG&OG&DG&DG\\
UDF1361&53.1652374&-27.8140640&19.75&4.38&3.14&2.15&2.90&0.57&0.153&0.70&-1.824&DG&DG&DG&OG&DG\\
\noalign{\smallskip}
\hline
\end{tabular}
}
\end{center}
Note. --- Column (4): $K$-band total magnitude in Vega; 
Columns (5) -- (6): the color of $i-K$ and $J-K$ in Vega; 
Columns (7) -- (8): the dust extinction and photometric redshifts 
of EROs; 
Columns (13) -- (16): classification results of EROs with different 
method, M1 from the SED fitting method, M2 from the $(i-K)$ vs. 
$(J-K)$ color diagram, M3 from the MIPS 24 $\mu$m image, M4 from the
nonparametric morphological indicators; Column (17): the final
classification results, considering the results of M1,M2, M3, and
M4.
\end{table}

In the right panel of Figure~\ref{fig:csed}, EROs in the UDF were
plotted in the $(i-K)$ vs. $(J-K)$ color diagram. 
The dashed line corresponds to the threshold of $(i-K) = 3.9$ for 
selecting EROs in this paper, the
dot-dashed line is a new color criterion for EROs classification,
based on evolutionary population synthesis model (see the next
subsection for detail). OGs, classified by the SED fitting method,
were plotted in red filled circles; DGs were plotted as blue open
circles. We found all OGs stayed at the left space of the dot-dashed
line, most (12/16) of DGs stayed at the right space of the
dot-dashed line.

\subsection{Classification based on $(i-K)$ vs. $(J-K)$
Diagram} \label{sec:ccd}

Pozzetti \& Mannucci.(~\cite{pozz00}) have introduced a method to
classify EROs into OGs and DGs based on their locations in the
$(I-K)$ vs. $(J-K)$ plane. EROs with $(J-K)>0.36(I-K)+0.46$ were 
classified as DGs, and the others are OGs. This method makes the 
$(I-K)$ vs. $(J-K)$ plane use a characteristic difference in 
the spectra of OGs and DGs located at $0.8 <z< 2$; OGs have a 
steep drop shortward of 4000 \AA, while DGs' spectra are smoother, 
giving DGs' $J-K$ colors redder than OGs.
As a consequence, OGs are located at the left part of the $(I-K)$ vs.
$(J-K)$ color diagram, while DGs are located at the right part.

\begin{figure*}
\centering
\includegraphics[angle=-90,width=0.85\textwidth]{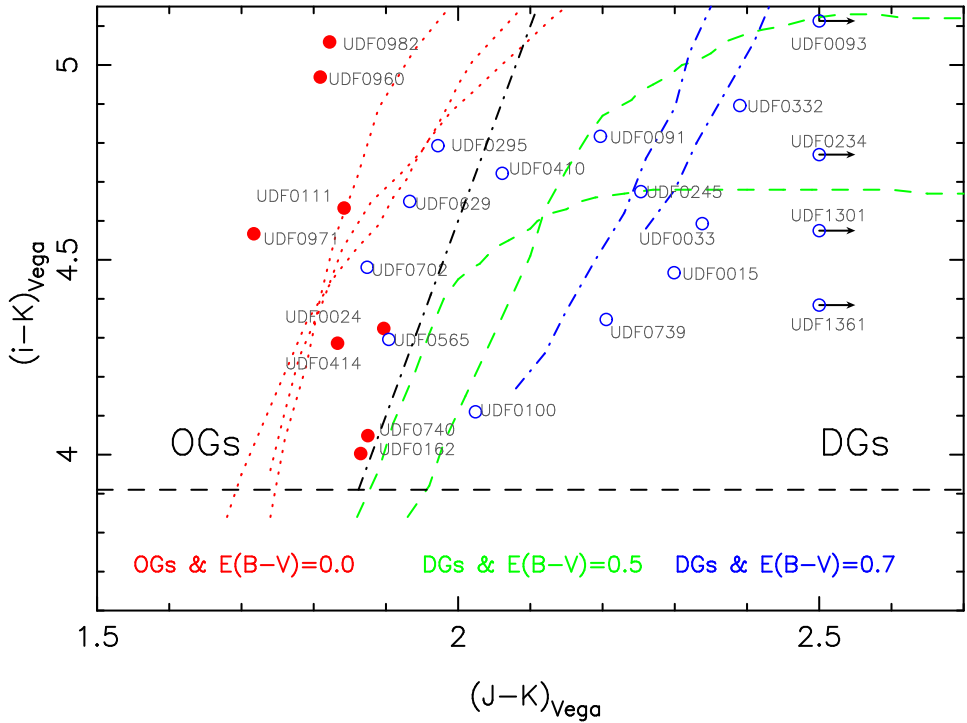}
\caption{ $(i-K)$ plotted against $(J-K)$
for 24 EROs in the UDF.  The dotted (elliptical galaxies with
$E(B-V)=0.0$, OGs), dashed (starburst galaxies with $E(B-V)=0.5$)
and dot-dashed (starburst galaxies with $E(B-V)=0.7$) lines show
color evolution of several representative model templates, using
KA97 models. The lines are plotted up to redshift of $z=2.5$.  The
thick dot-dashed line , $(J-K)=0.20(i-K) + 1.08$, 
corresponds to the boundary for separation of OGs and DGs
using $i-K$ vs. $J-K$ color.  The horizontal line and the data
points are the same as in Fig.~\ref{fig:csed}.}
\label{fig:cccd}
\end{figure*}

To classify the EROs in our sample, and check the validity of our
SED fitting method, we apply the color-color method to our sample
also.  Because the $i$-band (F775W) filter in the UDF is different
from the $I$-band filter used by Pozzetti \& Mannucci
.(~\cite{pozz00}), we have to develop our color criterion for EROs
classification. Figure~\ref{fig:cccd} shows $(i-K)$ vs.
$(J-K)$ model color-color diagram of
several representative galaxies with redshift $0.8<z<2.5$. Model
SEDs are adopted from the KA97 library, the lines represent
elliptical galaxies (pure bulge) with an extinction of $E(B-V)=0.0$
(dotted), starburst galaxies (pure disk) with $E(B-V)=0.5$ (dashed),
and starburst galaxies with $E(B-V)=0.7$ (dot-dashed), respectively.
From this figure, we found the color criterion 
$(J-K)=0.20(i-K)+1.08$ (thick dot-dashed line) can be used
to separate OGs and DGs for $0.8<z<2.5$. 
Using this new color criterion, we classified our EROs as OGs, if 
their $(J-K)<0.20(i-K)+1.08$. The others were classified as DGs.
The classification results based on the $(i-K)$ vs. $(J-K)$
color-color diagram are listed in the $14th$ column (M2, the second
classification method) of Table~\ref{tab:cls}.

In Figure~\ref{fig:cccd}, EROs are plotted as filled (OGs) and open
circles (DGs),  respectively, classified by the SED fitting method.
We found 12 EROs in the dusty starburst side, and the other 12 EROs
fell on the elliptical side of the division.  The agreement between
the SED fitting method and the $(I-K)$ vs. $(J-K)$ color diagram is
found to be satisfactory. All EROs classified as OGs by the SED
fitting method are also regarded as OGs by the $(i-K)$
vs. $(J-K)$ method. Similarly, 12 out of the 16 EROs
which are classified as DGs by the SED fitting method are located at
the dusty starburst side.  There are 4 EROs (UDF0295, UDF0565,
UDF0629 and UDF0702) which are classified as DGs by the SED fitting
method, but are located at the left-hand side of the thick
dot-dashed line. These galaxies are all close to the dividing line
and very faint in K-band, the photometric uncertainty of those faint
EROs may cause this discrepancy.

\subsection{Classification based on Spitzer MIPS 24 $\mu$m
Image}\label{sec:c24m}

Old stellar populations show a turndown at wavelengths longer than
the rest-frame 1.6 $\mu$m "bump", while dusty starburst populations
show emission from small hot dust grains and 6-12 $\mu$m polycyclic
aromatic hydrocarbon (PAH) features. Therefore, the difference of
ellipticals and dusty starbursting galaxies is very large at
mid-infrared, dusty EROs should have strong mid-infrared flux.
Between $z \sim 1- 2$, rest-frame 6-12 $\mu$m PAH and dust features
redshift into the 24 $\mu$m band. Any EROs detected at 24 $\mu$m
images should belong to the dusty population. Therefore, Spitzer
MIPS 24 $\mu$m data can be used to help us distinguish among
different ERO populations (Rieke et al.~\cite{riek04}; Werner et
al.~\cite{wern04}; Yan et al.~\cite{yan04}; Shi et
al.~\cite{shi06}).

\begin{figure*}
\centering
\includegraphics[angle=0,width=0.85\textwidth]{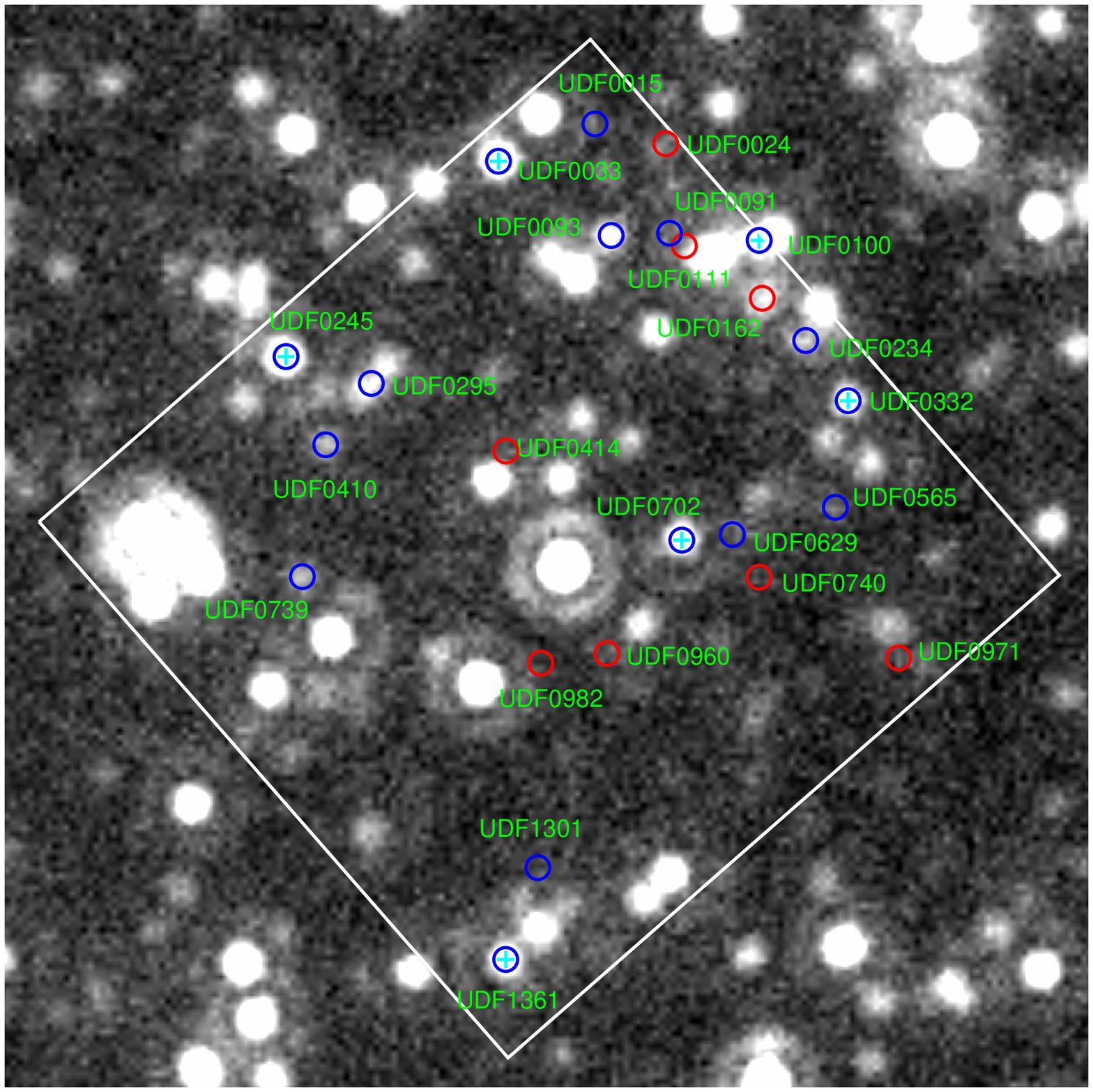}
\caption{ Distribution of the 24 EROs in the UDF on the Spitzer MIPS
24 $\mu$m image. Red circles and blue circles represent OGs and DGs,
respectively, classified by the SED fitting method. The size of
circle is in 2.4$''$ diameter.  EROs with $f_{24} > 80 $ $\mu$Jy are
plotted as cyan crosses. The white outline is same as in
Fig~\ref{fig:area}.} \label{fig:c24m}
\end{figure*}

A detailed description of the GOODS-S MIPS 24 $\mu$m observations,
data reductions, and data products can be found on Data Release 3
webpage. Source extraction at 24 $\mu$m was carried out using prior
positional information determined from the very deep IRAC 3.6 and
4.5 $\mu$m images, with a flux limit $\geq$ 80 $\mu$Jy. To merge our
ERO sample and the MIPS 24 $\mu$m source catalog, we used a simple
positional matching method with a 2.4$''$ match diameter, which
corresponds to a 3 $\sigma$ combined astronometric uncertainty from
the $K$-band and 24 $\mu$m data. However, of the 24 EROs in the UDF,
only 6 galaxies have 24 $\mu$m emission with flux greater than 80 
$\mu$Jy. We plotted these 6 EROs on the MIPS 24 $\mu$m image with
crosses (cyan) in Figure~\ref{fig:c24m}.

To check the reason for this low fraction of 24 $\mu$m-detected
EROs, we plotted all 24  EROs in the UDF on the MIPS 24 $\mu$m image
in Fig.~\ref{fig:c24m}.  EROs are plotted as red (OGs) and blue
(DGs) circles, respectively, classified by the SED fitting method.
We found that, beside these six 24 $\mu$m bright EROs, the other 8
EROs have counterparts in MIPS 24 $\mu$m image also. The reason that
these 8 EROs can not be found in the MIPS 24 $\mu$m catalog is
that the flux limit of the catalog, 80 $\mu$Jy, is too high for
faint EROs. We classify these $6+8$ EROs as DGs, and list them in
the $15th$ column (M3, the third classification method) of
Table~\ref{tab:cls}, the others as OGs by this method. For the 16
EROs in the UDF, which were classified as DGs by the SED fitting
method, 13 of them have counterparts on the the MIPS 24 $\mu$m
image; the left 3 EROs (UDF0091, UDF0629 and UDF1301) do not have
counterparts, and are very faint in $K$-band. For the 8 EROs in the
UDF, which were classified as OGs by the SED fitting method, only one
of them (UDF0162) has a counterpart on the MIPS 24 $\mu$m image, but
the distribution of 24 $\mu$m radiation around this ERO is very
diffuse.

\subsection{Classification based on Galaxy
Morphology}\label{sec:mor}

\begin{figure*}
\centering
\includegraphics[angle=0,width=0.85\textwidth]{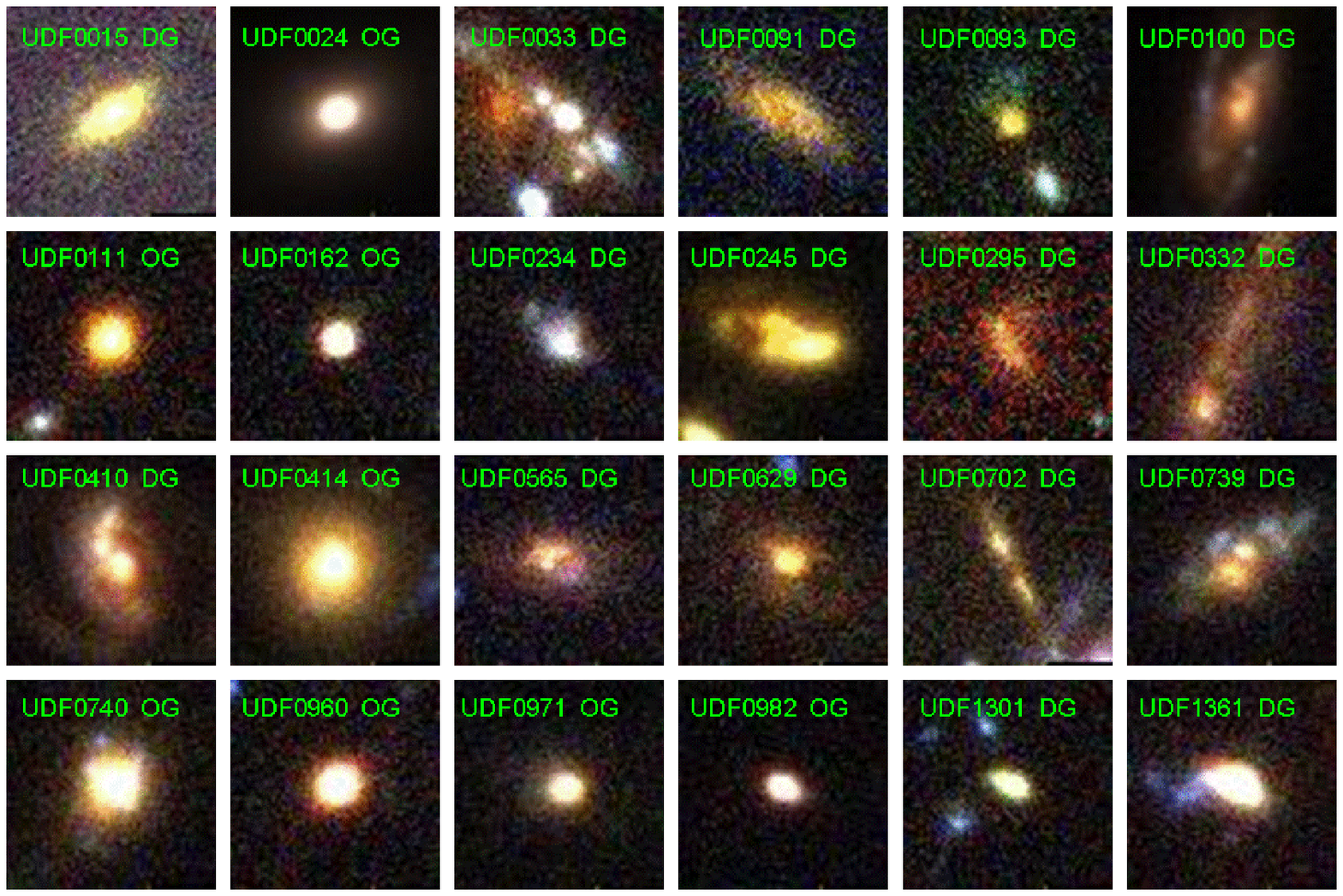}
\caption{ Color composite images of 24 EROs in the UDF. Red
represents the ACS$-z$ filter, green represents the ACS$-i$ filter,
and blue represents the ACS$-V$ filter. The regions shown are
$1.8\times1.8$ arcsec$^2$ in size, and north is down, east to the
right. Names and classification are displayed in upper right
corners. The classification here is based on the final 
classification results, considering the results of M1, M2, 
M3, M4.} \label{fig:ziv}
\end{figure*}

Figure~\ref{fig:ziv} shows the color images for 24 EROs in the UDF,
in which HST/ACS $z$-band, $i$-band and $V$-band were used as red,
green and blue color. These images have high spatial resolution
($0.03''$ pixel$^{-1}$), most of them show clear two dimensional
structures. To classify EROs as OGs or DGs, we have measured four
morphological parameters,  Gini coefficient (the relative
distribution of the galaxy pixel flux values, or $G$), \mt\, (the
second-order moment of the brightest 20\% of the galaxy's flux),
concentration index ($C$) and rotational asymmetry index ($A$) for
the EROs in our sample, with the high resolution ($0.03''$
pixel$^{-1}$) ACS-$i$ (F775W) images, and listed them in
Table~\ref{tab:cls}.

\begin{figure*}
\centering
\includegraphics[angle=-90,width=0.85\textwidth]{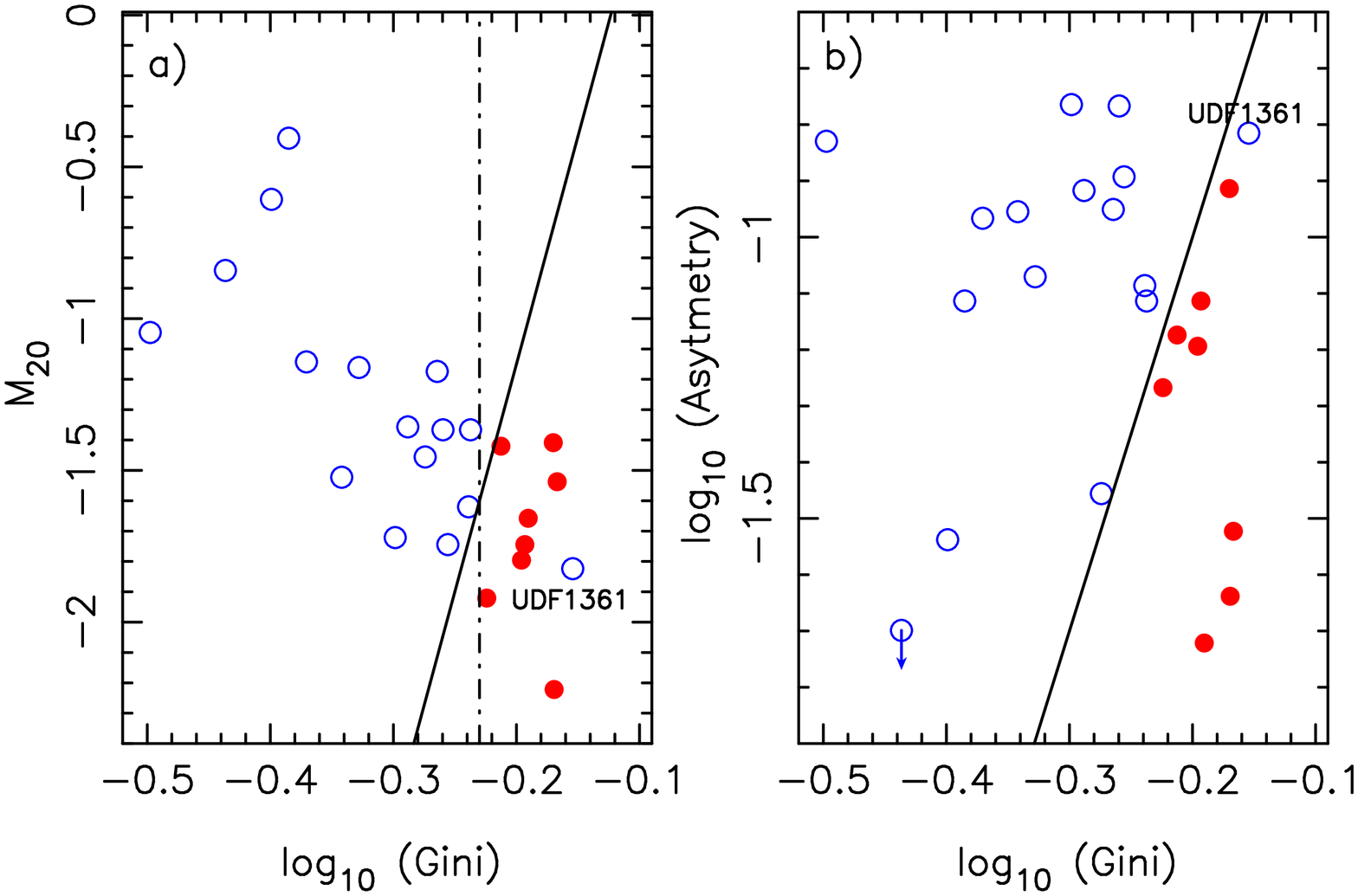}
\caption{ a) \mt\, versus Gini coefficient for EROs in the UDF. The
solid line is defined as $\mt = 15\log G + 1.85$, and the dot-dashed
line is defined as $\log G =-0.23$. b) Asymmetries versus Gini
coefficient.  The solid line is defined as $\log A = 7.0 \log G + 0.4$.
Filled- and open-circles represent OGs and DGs, respectively,
classified by the SED fitting method.} \label{fig:cmor}
\end{figure*}

The left panel of Figure~\ref{fig:cmor} shows the distribution of 24
EROs (filled and open circles represent OGs and DGs, classified by
the SED fitting method) in the log $G$ versus \mt\ plane.  The
distribution of EROs is very similar to that of local galaxies in
Lotz et al.(~\cite{lotz04}), with OGs showing high $G$ and low \mt\,
values, and DGs with lower $G$ and higher \mt\, values.  The solid
line is defined as $\mt = 15\log G + 1.85$, and the dot-dashed line
is defined as $\log G =-0.23$. Both of these lines can be used to
separate OGs and DGs, galaxies on the left side of them are DGs,
while on the right side are OGs, except for UDF1361.

The right panel of Figure~\ref{fig:cmor} shows the distribution of
EROs in the log $G$ versus log $A$ plane.  As found in Capak et
al.(~\cite{capa07}) for local galaxies, late type galaxies have
lower $G$ and higher $A$ values, early type galaxies have higher $G$
and lower $A$ values .  The solid line is defined as $\log A =
7.0\log G + 0.4$, can also be used to classify late type and early
type galaxies in our sample, except for UDF1361. As for UDF1361, it
was classified as a DG by the former three classification methods,
however, it was classified as OG by the morphology classification
method.  After checking its image in Figure~\ref{fig:ziv}, we found
that one of the possible reason is that the nonparametric
classification method can not separate ellipticals and early spiral
galaxies with a big bulge, it may also be because of recent or 
on-going merges and interaction of this galaxy.

Since the Gini coefficient has a very strong correlation with the
central concentration ($C$) for EROs in our sample,
and the relationship between $C-A$, $C-\mt$ are similar to those of
$G-A$, $G-\mt$, we do not plot these diagrams in this paper. As a
conclusion, we found that OGs have higher $G$ and lower \mt, $A$,
but DGs have lower $G$ and higher \mt, $A$; all of these structural
indices are efficient for separating OGs and DGs; the classification 
results (9 OGs and 15 DGs) are listed in the $16th$ column 
(M4, the fourth classification method) of Table 1. The classification 
of EROs, using morphological parameters, are in good agreement with 
the result based on the SED fitting method.

\section{Results and Discussion}

\subsection{Results}

As described in Section 3, we have developed four different methods 
to classify EROs into old passively evolving galaxies and dusty 
star-forming galaxies.
From Columns (13) -- (16) of Table~\ref{tab:cls}, we found that the 
agreement among these different methods is found to be satisfactory. 
For those 24 EROs in the UDF, 16 of them are classified as the same
ERO type (7 OGs and 9 DGs) by these different methods. 
However, for the other 8 EROs in our sample, they may be classified 
as OGs by one method, and while they may be classified as DGs by 
the other methods. 

For these four classification methods, the $(i-K)$ versus $(J-K)$ 
color-color diagram is simple, however, it depends on reddening, 
redshift, and photometric accuracy. Therefore, it is difficult to 
separate some objects of both classes fall near the discriminating 
line between starburst and elliptical. For 4 EROs 
(UDF0295, UDF0565, UDF0629 and UDF0702) 
in our sample, which are classified as DGs by the other 3 methods, 
but are located at the left-hand side of the 
discriminating line, and classified as OGs by the $(i-K)$ versus 
$(J-K)$ method. Considering these EROs are close to the dividing 
line and faint, the photometric uncertainty of them may cause this 
discrepancy. We classified them as OGs.

The Spitzer MIPS 24 $\mu$m image can help us to distinguish DGs 
accurately, by finding their counterparts in the mid-infrared band. 
However, due to the low spatial resolution of the Spitzer-MIPS 
instrument and low detection threshold, this method can not be 
used for very faint EROs. The classification results for the 3 
faint EROs (UDF0091, UDF0162, and UDF1031) by this method are 
different from the other 3 methods. We classify these 3 EROs into 
OGs and DGs based on the other 3 methods.

The SED fitting method and the nonparametric measures of galaxy 
morphology method almost offer the same result for EROs classifiction, 
except for UDF1361. UDF1361 was classified as DGs by the SED fitting 
method, the $(i-K)$ versus $(J-K)$ diagram, and the Spitzer MIPS 24 $\mu$m
image method, but the nonparametric measures of galaxy morphology 
method classified it as OGs. This EROs has late type SEDs, 
but elliptical type morphologies. We classify it as DGs.

We finally divide our EROs sample into 8 OGs and 16 DGs, 
corresponding to 33$\%$ and 67$\%$ of the whole sample. The detailed
results are shown in the last column of Table~\ref{tab:cls}. 
Although our sample is small, this ratio is consistent with the 
fractions given by previous works (Yan \& Thompson.~\cite{yan03}; 
Cimatti et al.~\cite{cima03}; Sawicki et al.~\cite{sawi05}). 
In other words, most of EROs (down to $K=22$) in our sample are DGs, 
which have spiral-like or irregular morphology.

\subsection{Discussion}

Galaxy morphology correlates with a range of physical properties in
galaxies, such as mass, luminosity, and, particularly, color, and this
suggests that morphology is crucial in our understanding of the
formation and evolution of galaxies (Li et al. 2007). Moreover, the
growing acceptance of the notion that galaxy morphology evolves
continuously throughout a galaxy's lifetime. However, because of
band-shifting effects, when we study the redshift evolution of
galaxy morphology, we have to use different rest-frame wavelength
images for comparison (for example, galaxies in the COSMOS field
have high spatial resolution images at HST/ACS F814W-band only).

To investigate the galaxy morphology as a function of wavelength, we
have measured the central concentration and the Gini coefficient for
 24 EROs in the UDF, using the deep and high spatial resolution
HST/ACS-$iz$ ($0.03''$ pixel$^{-1}$) and HST/NICMOS-$JH$ ($0.09''$ 
pixel$^{-1}$) band images.
The results are listed in Table~\ref{tab:ijh}. Column (1) lists the
galaxy name. Columns (2) -- (9) list the morphological parameters of
$G$ and $C$, using the HST $i$-, $z-$, $J$-, and $H$-band images.
The mean values of these  morphological parameters are listed in the
last two rows of Table~\ref{tab:ijh} for OGs and DGs, respectively.

\begin{table}
\setlength{\tabcolsep}{1.1mm}
\begin{center}
\caption[]{Gini coefficient and concentration index at HST/ACS
$i$-band (F775W), $z$-band (F850LP), HST/NICMOS $J$-band (F110W) 
and $H$-band (F160W) for the 24 EROs in the UDF.\label{tab:ijh}}
\begin{tabular}{lrrrrrrrrrrrrrr}
\noalign{\smallskip}
\hline
\hline
\noalign{\smallskip}
NAME    &$G$(F775W)&$C$(F775W)&$G$(F850LP)&$C$(F850LP)&
$G$(F110W)&$C$(F110W)&$G$(F160W)&$C$(F160W)\\
(1)&(2)&(3)&(4)&(5)&(6)&(7)&(8)&(9)\\
\hline\noalign{\smallskip}
        &\multicolumn{8}{c}{OGs}\\
\cline{2-9}\noalign{\smallskip}
UDF0024&0.677&0.576&0.680&0.581&0.700&0.562&0.703&0.590\\
UDF0111&0.645&0.491&0.662&0.458&0.653&0.492&0.698&0.527\\
UDF0162&0.676&0.524&0.646&0.479&0.498&0.346&0.508&0.350\\
UDF0414&0.597&0.449&0.614&0.460&0.640&0.504&0.695&0.565\\
UDF0740&0.637&0.497&0.664&0.504&0.653&0.484&0.715&0.555\\
UDF0960&0.681&0.486&0.700&0.505&0.682&0.452&0.755&0.562\\
UDF0971&0.641&0.503&0.646&0.493&0.677&0.536&0.714&0.566\\
UDF0982&0.603&0.355&0.616&0.428&0.611&0.349&0.699&0.545\\
\noalign{\smallskip}\hline
\\
        &\multicolumn{8}{c}{DGs}\\
\cline{2-9}\noalign{\smallskip}
UDF0015&0.532&0.357&0.566&0.399&0.483&0.298&0.580&0.348\\
UDF0033&0.503&0.329&0.443&0.225&0.417&0.262&0.567&0.409\\
UDF0091&0.366&0.171&0.375&0.193&0.441&0.248&0.556&0.349\\
UDF0093&0.555&0.413&0.585&0.422&0.581&0.378&0.689&0.533\\
UDF0100&0.515&0.307&0.542&0.340&0.432&0.255&0.518&0.336\\
UDF0234&0.455&0.325&0.435&0.277&0.410&0.276&0.525&0.297\\
UDF0245&0.550&0.368&0.581&0.346&0.559&0.366&0.653&0.440\\
UDF0295&0.399&0.214&0.490&0.209&0.426&0.210&0.516&0.311\\
UDF0332&0.412&0.179&0.295&0.131&0.303&0.202&0.445&0.239\\
UDF0410&0.544&0.306&0.502&0.324&0.577&0.394&0.638&0.481\\
UDF0565&0.470&0.208&0.494&0.285&0.487&0.311&0.558&0.353\\
UDF0629&0.577&0.383&0.587&0.396&0.542&0.428&0.635&0.501\\
UDF0702&0.426&0.231&0.450&0.229&0.545&0.374&0.581&0.347\\
UDF0739&0.318&0.137&0.445&0.269&0.545&0.346&0.636&0.425\\
UDF1301&0.579&0.308&0.542&0.289&0.492&0.253&0.662&0.517\\
UDF1361&0.701&0.569&0.647&0.475&0.650&0.502&0.741&0.635\\
\noalign{\smallskip}\hline
\\
Mean-OGs&0.645&0.485&0.654&0.489&0.639&0.466&0.686&0.533\\
Mean-DGs&0.494&0.300&0.499&0.301&0.493&0.319&0.594&0.408\\
\noalign{\smallskip}\hline
\end{tabular}
\end{center}
\end{table}

\begin{figure*}
\centering
\includegraphics[angle=-90,width=0.85\textwidth]{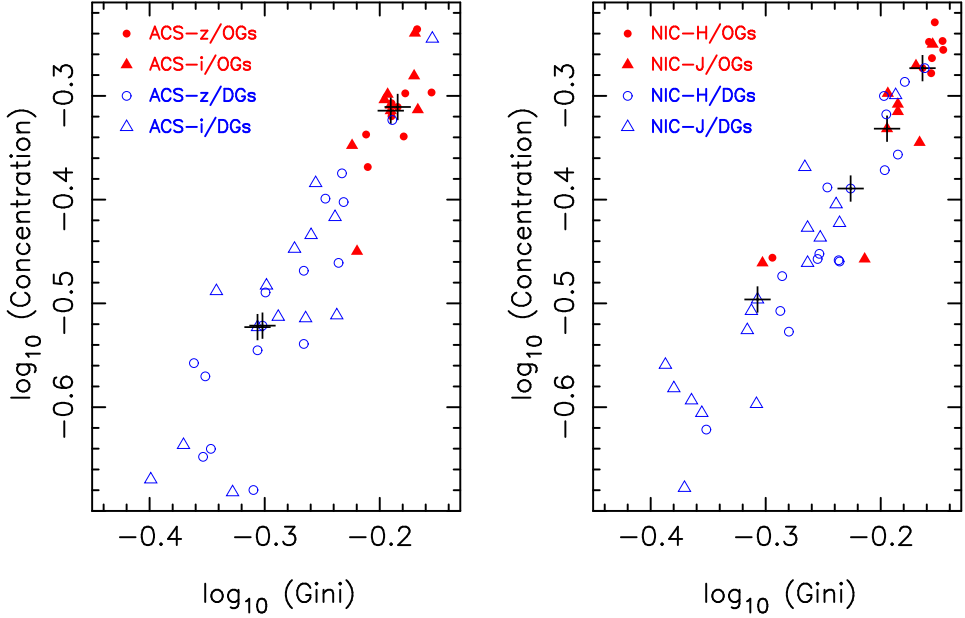}
\caption{Gini coefficient versus concentration index for EROs in the
UDF. Left panel shows morphological parameters of HST/ACS $i$- and 
$z$-band, which have spatial resolution at $0.03''$ pixel$^{-1}$;
Right panel shows morphological parameters of HST/NICMOS $J$- and 
$H$-band, which have spatial resolution at $0.09''$ pixel$^{-1}$.
OGs are shown as red symbols, and DGs as blue. The mean values for DGs 
and OGs in different bands are shown as pluses.} \label{fig:mgc}
\end{figure*}

Figure~\ref{fig:mgc} shows the distribution of EROs in the log $G$
versus log $C$ plane. The EROs classified as OGs in the previous
section are shown as red, DGs as blue. 
Firstly, a strong correlation between the Gini coefficient and the
concentration index can be found for $z\sim1$ EROs, as found by
Abraham et al. (2003) for local galaxies. The correlation between
$C$ and $G$ exists because highly concentrated galaxies have much of
their light in a small number of pixels, then have high $G$ values.
Secondly, OGs have higher $G$ and $C$ values than DGs. This is what
we have expected, since DGs are known to contain star formation and
asymmetries produced by star formation or merging and tidal
interactions with other galaxies. These galaxies thus have lower
central concentration. 
Finally, we can find that the morphological parameters of galaxies 
in our sample depend on the wavelength of observation, from 
Figure~\ref{fig:mgc} and Table~\ref{tab:ijh}. 
To show this point more clearly, we plot the mean values of $G$ and 
$C$ (pluses) for both OGs and DGs at each band as in 
Figure~\ref{fig:mgc}, and list the mean values in the last two rows
of Table~\ref{tab:ijh}. 
For the same spatial resolution imaging data sets, galaxies have 
higher $G$ and $C$ when these morphological parameters are measured 
at longer wavelength bands, the similar findings can also be found 
in Abraham et al. (2003).
Therefore, when studying the morphological evolution of galaxies as a
function of redshift, morphological parameters should be measured 
with the same (or similar) rest-frame wavelength images.

\section{Summary}

We have described the construction of a sample of extremely red
objects within the Hubble Ultra Deep Field images, and developed
four different methods for their classification. Taking advantage
of the high-resolution HST/ACS and HST/NICMOS imaging, we also
analyzed morphological parameters of galaxies while considering 
wavelength. Our main conclusions are as follows:

(1) We identify a sample of 24 EROs, defined here as $(i-K)> 3.9$ 
galaxies, to a limit $\Kv=22$ in the Hubble Ultra Deep Field.  
Compared to OGs, we find that most of the DGs are in the range of 
faint magnitude, while the fraction of OGs is similar to that 
of DGs to \Kv $ \leq $ 20.5.

(2) To classify EROs in our sample, we develop four different
methods, the SED fitting, $(i-K)$ vs. $(J-K)$ color, MIPS 24 $\mu$m
image, and nonparametric measures of galaxy morphology. We found
that the classification results from these methods agree well.

(3) Combining these methods, we separate OGs and DGs in our EROs
sample. About 33$\%$ and 67$\%$ of them are classified as OGs and
DGs, respectively. 

(4) We measure the morphological parameters of $G$ and $C$ for the
24 EROs in the HST/ACS $i$-, $z$- ($0.03''$ pixel$^{-1}$) and 
HST/NICMOS $J$-, $H$-band ($0.09''$ pixel$^{-1}$) images in the UDF, 
respectively. For the same spatial resolution data sets, the results 
show that both OGs and DGs have lower $G$ and $C$ values at shorter 
wavelength bands.
Furthermore, the strong correlation between the Gini coefficient
and the concentration index of galaxies can be found at each band.

Considering that our EROs sample is small, we plan to use these
classification methods to much larger fields, such as GOODS, GEMS
and COSMOS, etc, examining the efficiency of these methods and
studying the classification and physical properties of EROs in the
future.

\begin{acknowledgements}
We would like to thank Simon D. M. White and Emanuele Daddi for
their valuable suggestions.  We are grateful to Robert G. Abraham
and Christopher J. Conselice for their comments and access to their
morphology analysis codes.  The work is supported by the National
Natural Science Foundation of China (NSFC, Nos. 10573014,
10633020, and 10873012), the Knowledge Innovation Program of the
Chinese Academy of Science (No. KJCX2-YW-T05), and National Basic
Research Program of China (973 Program) (No. 2007CB815404).
\end{acknowledgements}

\label{lastpage}
\end{document}